\pdfoutput=1
\documentclass[aps]{article}
\usepackage{geometry}                
\usepackage{graphicx}

\usepackage{amssymb}
\usepackage{epstopdf}
\usepackage{amsmath}

\usepackage{amsfonts}
\usepackage{bm}
\usepackage{hyperref}

\usepackage{tikz-cd}
\usepackage{physics}

\DeclareMathAlphabet{\mathpzc}{OT1}{pzc}{m}{it}
\newcommand{\sgn}{\operatorname{sgn}}

\date{October 26, 2018}
\title{Essential Infinite Order Non-PDE Behavior in Continuum Mechanics: Corrections to Hydrodynamics and Diffusion}
\author{Clifford Chafin\\\ \small{Department of Physics, North Carolina State University, Raleigh, NC 27695} \thanks{cechafin@ncsu.edu}}

\begin{document}
	\maketitle
	\begin{abstract}
	Longstanding problems regarding the causality of the diffusion equation are resolved through a class of exact solutions.  A universal differential solution for diffusive processes is derived that is causal and exact at any analytic point in the data, albeit infinite order in spatial derivatives.  This is true for systems both relativistic and nonrelativistic and shows that the hyperbolic and other relativistic extensions of the heat equation are not valid.  A similar effect is demonstrated for flow enhanced mixing of solutions but with a new nonanalytic feature.  
	Viscous hydrodynamics of liquids have both features.  Both Newtonian and non-Newtonian viscous liquids give a more important and confounding alteration of the N-S equations for nonstationary flows.  A careful analysis of liquids in terms of microscopic constituents and Lagrangian paths show there is a well-defined unique microscopic decomposition of fluid deformations into rotation and two modes corresponding to organized layered flow that do not enhance mixing and a mode that induces mixing and preferred static orientations.  The resulting equations are both infinite order in spatial derivatives almost everywhere and are divided into two disjoint classes by an essential nonanalytic hypersuface.  These give important rheological effects coupling hydrodynamic flow to diffusion and reaction kinetics.  Practical consequences include catalysis and reaction yield control by rheological means.  Implications of this work should percolate through almost all of continuum mechanics.  
	\end{abstract}
\section{Introduction}
The history of continuum mechanics of material media has been closely tied up with the development of partial differential equations and their solutions and dates back to a time long before the atomistic picture of matter was popular and accepted.  Hydrodynamics and elasticity were the first physical motivation and testing grounds for these ideas.  These ideas were then exported to electromagnetism where the fields were often thought of as fluid flows and the waves as propagation through a very light and stiff medium: the luminiferous aether.  Later the aether was abandoned and Einstein showed it was better to modify mechanics and that electromagnetism was already perfect as it stood.  Here I will demonstrate that, while continuum mechanics motivated the theory of PDEs, which is perfect for electromagnetism, general relativity and the theories of the fundamental fields of classical physics, these higher level theories of collections of particles typically have dynamics too general to be described by the mathematical theory they motivated.  

In the case of diffusion and heat conduction, causality has been long recognized as a problem.  Fick's law and Fourier's law are the essentially equivalent equations we learn as undergraduates for such physics.  Not only does the propagation front from these laws travel faster than the characteristic speed of the medium (thermal velocity, sound, etc.) but faster than light itself.  This generally results in only small effects but it is still problematic.  Popular fixes are the hyperbolic heat conduction (HHC) equation, the relativistic heat conduction equation (RHC) and appeals to the porous medium equation \cite{Molina}.  Objections are the lack of physical motivation of parameters, violation of the second law and, as I personally emphasize, they introduce kinematic degrees of freedom not consistent with the microscopic physics where only position and velocity of particles can be freely specified.  Resolution of this is found in a careful analysis of the discrete scales of the system that do not vanish in some physically available limit.  When pushed into a best possible partial differential relation, this results in an equation that is first order in time but infinite order in the spatial derivatives.  

Mixing is generally thought of as being driven quickly by convection then locally by diffusion but I will show that certain kinds of flows force an irreversible mixing directly and this leads to a similar infinite order set of equations.  Furthermore, there is an essential nonanalytic feature to these equations, unlike in the diffusive case.  

In the case of hydrodynamics, the problems are even more profound.  The causality problem of diffusion of momentum is inherited but, beyond that, in the inviscid limit Navier-Stokes is accurate.  A novel derivation of these equations are given that is purely Lagrangian, based on particle trajectories, is given.  Unfortunately, as is known, the limit of hydro for a simple viscous fluid as viscosity vanishes does not give such Eulerian behavior since turbulence grows without bound in this limit.  This Eulerian limit is best for transient flows with low drag and leading edge flows where turbulence has not had time to arise.  When we introduce viscosity, I will show that the usual appeal to the deviator to build the internal forces is not sufficient for time changing flows.  Correct treatment of this leads to a differential relation in the hydrodynamic parameters that becomes infinite order when converted into a differential equation in the flow velocity.  This should be of substantial interest to those pursuing the Millennium Challenge problem for the long time solutions and stability of the Navier-Stokes equations.  A very nice benefit of this approach is that it gives a unique physical way to separate the local flow into a rotational component and two other local flows that are microscopically relevant, something not possible using the purely linearization of the local velocity field.  That has important implications for both rheology and its effects on chemical reactions.

\section{Diffusion}
\subsection{1-D Diffusion}
To address the problem of causality and diffusion we return to the fundamental microscopic understanding of the effect.  Diffusion often refers to heat and particle migration.  Particles migrate through a liquid by kinetic fluctuations, through gases via kinetic ballistic motion and heat migrates in condensed matter by phonon migration interrupted by imperfections in lattices and thermal fluctuations in its underlying order.  I will consider the simplified model of particles that move by $\Delta x$ along a line at a constant speed of $v$ whereupon it is reflected with equal likelihood right or left with the same velocity. Later, I will argue that such models can be combined to describe any more complicated diffusive process.  This gives a collision time of $\Delta t=\Delta x/v$.  The governing dynamics can then be described by 
\begin{align}\label{numberdiff}
n(x,t+\Delta t)=\frac{1}{2}\left(n(x-\Delta x,t)+n(x+\Delta x) \right)
\end{align}
where $n$ is the number of particles at $(x,t)$.  
Note that this is evidently causal with the edges of the support moving outwards at the characteristic speed $v$.  Additionally the system partitions itself into disjoint copies separated by some fraction of unity so that we actually have a continuum of disjoint systems parameterized by $[0,1)$.  Of course, we are biased against such models for several reasons not the least of which is that it is a finite difference problem rather than a pde.  Further, we would like our data to be described in terms of a density function, $\rho(x,t)$ rather than a discrete number $n(x,t)$.  If we enforce that the initial data is smooth to all orders then we can rewrite eqn.\ref{numberdiff} by replacing $n\rightarrow \rho$ and then using a Taylor expansion to rewrite it in terms of derivatives which gives
\begin{align}\label{rho}
	\rho(x,t)+\partial_t \rho(x,t)\Delta t+\frac{\partial_t^2 \rho(x,t)}{2!}\Delta t^2 + \ldots=\rho(x,t)+\frac{\partial_x^2 \rho(x,t)\Delta x^2}{2!}+\frac{\partial_x^4 \rho(x,t)\Delta x^4}{4!}+\ldots
\end{align}
We expect the usual kinematic behavior and so attempt a first order in time differential solution.  It is fairly evident that no finite order differential equation will arise so I search for a differential equation of the form
\begin{align}\label{form}
	\partial_t\rho(x,t)=\sum_{n=1}^{\infty} a_n \partial_x^n \rho(x,t).
\end{align}
Iterating eqn.\ref{form} in eqn.\ref{rho} we get the LHS of eqn.\ref{form}
\begin{align}\label{lhsform}
	\rho(x,t)+\sum\limits_{n=1}^{\infty}a_n\partial_x^n \rho(x,t) \Delta t+\frac{\sum\limits_{m=1}^{\infty}\sum\limits_{n=1}^{\infty}a_m a_n\partial_x^{m+n} \rho(x,t)}{2!}\Delta t^2\\ + \frac{\sum\limits_{l=1}^{\infty}\sum\limits_{m=1}^{\infty}\sum\limits_{n=1}^{\infty}a_l a_m a_n\partial_x^{l+m+n} \rho(x,t)}{3!}\Delta t^3 \ldots
\end{align}
The parameters $\Delta x$ and $\Delta t$ are not used as ``small" parameters here for an expansion, rather we solve by matching derivative orders, $\partial_x^n \rho$, and finding the coefficients of eqn.\ref{form}.  The only nonzero coefficients are even and we obtain a set of recursively solvable equations (for even N) 
\begin{align}
	a_N  \Delta t+\sum_{k=1}^{N-1}a_{N-k}a_k \Delta t^2 +\sum_{k=1}^{N-2}\sum_{l=1}^{N-k-1}a_{N-k-l}a_k a_l\Delta t^3+\ldots +a_1^N \Delta t^N = \frac{\Delta x^N}{N!}
\end{align}
where we understand all $a_k=0$ when $k$ is odd (so the last term is illustrative but extraneous).  
This gives the following relationship in the form of an infinite series in derivatives for diffusion:
\begin{align}\label{diffusioneom}
	\partial_t \rho&=v\left( \frac{1}{2}\Delta x\partial_x^2\rho -\frac{5 }{24} \Delta x^3 \partial_x^4\rho + \frac{61}{720}\Delta x^5  \partial_x^6 \rho -\frac{277}{8064}\Delta x^7 \partial_x^8\rho + \frac{50521}{3628800}\Delta x^9 \partial_x^{10} \rho + \ldots\right)\\
	&=-\frac{v}{\Delta x}\left( \sec(i\Delta x \partial_x)-1\right)\rho(x,t)\\
	&=-\frac{v}{\Delta x}\left( \sech(\Delta x \partial_x)-1\right)\rho(x,t)\\
	&=\partial_x\Big[-v\hat{O}(\Delta x \partial_{x}) \rho(x)\Big]\\
	&=\partial_x j(x)
\end{align}
where
\begin{align}
\hat{O}=\frac{\sech(u)-1}{u}.
\end{align}
This is clearly a modification of the usual diffusion equation with $\frac{1}{2}v\Delta x$ playing the role of the diffusion constant. If we let $v\Delta x$ stay finite as $\Delta x\rightarrow 0$ we have the usual Fick/Fourier diffusion pde and see that $v\rightarrow\infty$ so we expect the advancing front of a distribution with compact support to move with infinite speed. Causality is violated because the underlying processes move at an infinite speed which also governs the front speed.  
 
Since this equation is essentially infinite in order it should be thought of as a ``limit pde.''  Often infinite order equations are shunned as ``nonlocal'' but this is a little unfair.  The temporal order is only one so the initial data is adequately given by the value at a single time.  The underlying physics here is of discrete finite processes and we are forcing continuum shoes on these discrete feet.  Even so, these derivatives are all local information to derive the time variation of the density at each point. Later we will see that hydro requires such a limit pde expansion to be put in a familiar form where no nonlocal physics is at work.  In general, I believe that integral equations with spatially compact kernels will also give causally restricted behavior and be able to be put into such a form.  It would be especially interesting if the converse, causally evolving linear limit pdes, correspond to some general compact kernel of an integral equation.  

It is a good moment to reflect on why this result deviated from the usual derivations of diffusion or transport based on the usual conserved quantities.  In these derivations the current, $j$, is defined as $n\cdot v$ where $v$ is the flow velocity or one uses some first order parcel to derived flux across the ends.  Such derivations assume that the current is of low finite order.  This is always true for the fundamental fields of nature but for macroscopic and classical media where there is significant quantum localization present, we see that this is not necessarily true.  Even though such a model may give the lowest order contribution correctly, there can be infinite higher order contributions, as we see here.  

Notice that the free parameters in this model are $(v,\Delta x)$ and the equations are linear.  If we had many different processes going on as in the case of a distribution of velocities and step sizes as might be encountered in the Boltzmann description of a gas or richer description of diffusion we can combine these in the following way
\begin{align}\label{distribution}
	\partial_t \rho&=\int_0^\infty \int_0^\infty dv d(\Delta x) h(v,\Delta t) v \sum\limits_{n=1}^{\infty} a_n(\Delta x) \partial_x^n \rho(x,t)\\
	&=\sum\limits_{n=2}^{\infty}\partial_x^n \rho(x,t) \cdot \int_0^\infty \int_0^\infty dv d(\Delta x)  h(v,\Delta t) v a_n(\Delta x) \\
	&=\sum\limits_{n=2}^{\infty} F(n) \partial_x^n \rho(x,t)
\end{align}
where $h(v,\Delta x)$ is a normalized distribution function over the parameter space of velocity and step size and the $a_n(\Delta x)$ set is the universal coefficient function set from eqn.\ref{diffusioneom}.  This distribution function can have built in velocity limits associated with relativity or the medium properties.  A population of random walkers that is randomly traversing a velocity distribution has each walker eventually return arbitrarily closely to its original velocity.  To the extent the distribution of velocity and step size is stable over the support of the spatial density profile, we can view these as effectively independent distributions which would justify eqn.\ref{distribution}.  The advancing edges of the support will tend to be advanced by rarer cases of higher velocity and larger step size which then may take a relatively long time to equilibrate through the various $(v,\Delta x)$ populations.  This may lead to a weakening of the validity of this form but, for most diffusive applications, this will be a small effect and can only modify the dynamics in such a way as to further thin the growing edge but not alter its speed from $v$.  It would be interesting to know if the full dynamics which include the advancing thin edge of the distribution can be modeled by any linear limit pde at all.  Given a distribution of velocity and step sizes it seems likely that nonlinearity will take over as the mixing between different populations becomes altered based on the local distribution.  
In the case of hydrodynamic flow with diffusion, especially relativistic flow, we will need to first reconsider the foundation of hydrodynamics below, which will raise many profound issues before such an approach can be formulated.

\subsection{Higher Dimensional Diffusion}
To generalize the previous section we need to know something about the isotropy of the medium.  If it is spherically symmetric then we need to look at an N-sphere about the point at distance $\Delta x$ where particles at velocity $v$ travel in to form the new density at a time $\Delta t$ in the future.  This introduces a whole continuum of variation in density that we must include to all orders in constructing an analogous equation whereas, in the 1D case, we only needed two points!  For a 2-sphere (a circle) we have a flux rate into a differential angle $d\theta$ that reaches the receiving location at $\Delta x$ from each point.  The flux radiated from a single point to the circle must equal the net flux radiated from a single point so we obtain
\begin{align}\label{rho2D}
	\rho(x,t+\Delta t)=\frac{1}{2\pi}\int_0^{2\pi} \rho(x+\Delta x\cdot \hat{r}(\theta),t)d\theta
\end{align}
If we have a spherically symmetric distribution then this will be preserved in time.  Therefore a solution analogous to eqn.\ref{form} must have the form
\begin{align}\label{form2D}
\partial_t \rho(x,t)=a_2 \nabla^2\rho+a_4 (\nabla^2)^2\rho+\ldots
\end{align}
Similarly this implies that the integrations will pull out the spherically symmetric parts of the sums in eqn.\ref{rho2D}	
\begin{equation}
\begin{split}
\rho(x,t)+\partial_t &\rho(x,t)\Delta t+\frac{\partial_t^2 \rho(x,t)}{2!}\Delta t^2 + \ldots\\
&=\rho(x,t)+\frac{1}{2\pi}\int_0^{2\pi}\left(\frac{\sum\limits_{ij} r_{i}r_{j}\partial_{ij}\rho(x,t)\Delta x^2}{2!}+\frac{\sum\limits_{ijkl} r_{i}r_{j}r_{k}r_{l}\partial_{ijkl}\rho(x,t)\Delta x^4}{4!}+\ldots\right)d\theta\\
&=\rho(x,t)+\frac{\nabla^2\rho(x,t)\Delta x^2}{2!}+\frac{(\nabla^2)^2\rho(x,t)\Delta x^4}{4!}+\ldots
\end{split}
\end{equation}
where $r_i(\theta)$ is the index representation of the unit vector $\hat{r}$ pointing in the direction of the angle $\theta$.  Recognizing that only the even terms are nonzero in eqn.\ref{lhsform} we see that solving for the $a_n$ parameters in eqn.\ref{form2D} gives exactly the same equations in $\nabla^2\rho$ as for $\partial_x \rho$ so that the expansion for eqn.\ref{diffusioneom} is exactly the same with the replacement $\partial_x^2\rightarrow\nabla^2$, $\partial_x^4\rightarrow(\nabla^2)^2$, etc.  This argument is easily generalized to higher dimensions and shows that the coefficient functions, $a_n(\Delta x)$ are universal, i.e.\ independent of dimension.  

\section{Extension Flow Driven Mixing}\label{mixing}
\begin{figure}
	\begin{center}
		\includegraphics[trim = 0mm 0mm 10mm 10mm, clip,width=3cm]{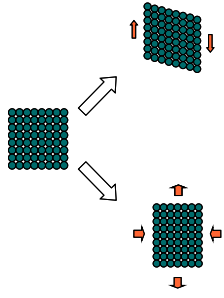}
		\caption{A lattice structure under shear and extension deformations.}
		\label{fig:shearvext}
	\end{center}
\end{figure}

\begin{figure}
	\begin{center}
		\includegraphics[trim = 10mm 20mm 10mm 0mm, clip, height=3cm]{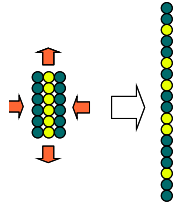}
		\caption{Mixing rearrangements under extension deformation.}
		\label{fig:extmix}
	\end{center}
\end{figure}

\begin{figure}
	\begin{center}
		\includegraphics[trim = 10mm 0mm 0mm 0mm, clip, height=3cm]{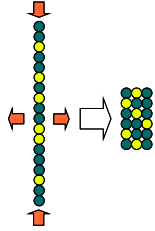}
		\caption{The fluid parcel after the reversed extension deformation.}
		\label{fig:extmixrev}
	\end{center}
\end{figure}

Diffusion is generally a mixing of energy or particles by thermal means and induced fluctuation driven rearrangements.  Convective mixing drives fluid into swirling parcels and plumes that move heat and material about in a coarse grained fashion.  Once strong gradients are locally formed, diffusion finishes the process of driving uniformity.  There is, however, a means for flow to drive mixing on the microscopic scale.  When pure laminar shear flow occurs, like Couette flow, particles tend to slide past each other as in fig.\ref{fig:shearvext} but during extension, particles must randomly pass each other in an interleaving pattern along the lengthening axis as in fig.\ref{fig:extmix}.  Conversely, when extending in the perpendicular direction, the particle randomness acquired is preserved and no density changes along the original axis of mixing are created as in fig.\ref{fig:extmixrev}.


An application of this notion is in the effect on concentration gradients in solutions of pure extension flow.  For a pair of parcels at the center of such an extension flow $\vec{v}=(-x,y)$ with a concentration function $\rho(x)$ (so that there is no $y$-variation in concentration) we can find the effect of a two-fold change in the parcel by averaging over the pair of $\Delta s$ thickness layers.  This happens over the time scale set by $u=\Delta s \nabla \vec{v}$ so $\Delta t=\Delta s/u$.  This gives an equation identical to eqn.\ref{numberdiff}.  At other points there is advective translation of the parcels in a global frame so we need to include this but locally the same discrete process is at work.  In contrast, the reversed flow $\vec{v}=(x,-y)$ does no mixing at all.  This gives the following advective unidirectional diffusion equation for concentration gradient $\rho(x)$ for an extension flow aligned with the coordinate axes, $\vec{v}=(-ax,ay)$,
\begin{align}
	\partial_t\rho(x,t)+\vec{v}\cdot\nabla \rho(x,t)=\begin{cases}
	-a\left( \sech(\Delta x \partial_x)\rho(x,t)-1\right)        & \text{if } a \geq 0 \\
	0        & \text{otherwise}
	\end{cases}
\end{align}
When I later investigate a physically meaningful unique decomposition of hydrodynamic flow, the distinction of uniform Couette type shear with the interleaving behavior of extension and its associated effect on concentration gradients will be one of the motivating features.  This is a second example of essentially infinite order continuum behavior and, furthermore, one with a nonanalytic partition of domains in its behavior.

\section{Hydrodynamics}
In gases, we have reason to worry that Navier-Stokes will have infinite order corrections because momentum diffuses in them the same way particles do: diffusively.  In liquids, the molecules are in close contact with transient bonds and so can maintain an elastic shear stress, as solids do but now the stress is only from dynamic relative changes rather than static ones.  The length scales in the limit of granularity are less interesting so we might expect that N-S would be better in this respect.  Most derivations of the N-S equations are built on conservation law considerations or by directly appealing to parcel motion\cite{Batchelor}.  Given the previous sections' results, one might worry that the deformation of the parcels in simple Newtonian derivations of the equation of motion might introduce some small higher order correction that we have missed.  There are no intrinsic length and time scales that would seem to give an expansion like we saw for diffusion with discrete steps but one can certainly build length and time scales from ratios of derivatives of the velocity and density field e.g.\ $\tau=(\hat{v}\nabla v)^{-1} $ so it is, in principle, possible that an infinite order expression could arise even in the case of inviscid liquids.  
N-S is rarely (if ever) derived directly and unambiguously from Newton's laws (i.e.\ using particles instead of parcels) so let us do so here and verify that, in this case, they are indeed valid.  
\subsection{1-D Hydrodynamics}

\begin{figure}[h]
	\centering
	\includegraphics[width=2in,clip]{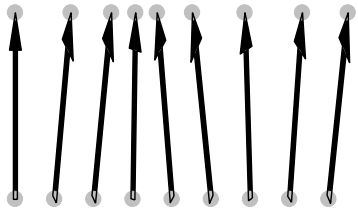}
	\caption{Map from unifrom index space to location space $u\rightarrow x$. }
	\label{fig:map}
\end{figure}

To derive an ``Eulerian" form of inviscid hydrodynamics from the Lagrangian picture (or maybe we should say Newtonian picture since every particle has a coordinate label and is moving according to Newton's laws) we consider a quasi-continuum of particles indexed by the set $u\in\mathbb{R}$.  We can think of the coordinate of the $u^{th}$ particle as $s(u)$ as originating from a uniform distribution where $s(u)=u$ that has now been deformed.  Each particle has an (infinitesimal) mass $dm$.  To maintain a single valued velocity to the flow there needs to be some pressure type forces that prevent particles from ever overlapping.  Given this we can assume that a smooth mapping $u\rightarrow s$ that persists for all time as in fig.\ref{fig:map}.  

Since $N=\int n(x)dx \sim \int du$, the density of particles at $x=s$ is given by 
\begin{align}\label{density}
	n(x)=\alpha (ds/du)^{-1}|_{u=s^{-1}(x)}
\end{align}
 where $\alpha$ is a real constant.

The equation of motion for each particle is $F=ma$ which for our particles becomes
\begin{align}\label{eom}
	\Delta m\frac{\partial^2}{\partial t^2}s(u,t)=-\Delta P(x,t)
\end{align}
where $P(x)$ is given by the x-density of particles, $P(x)=\beta n(x)$.  This is clearly not an equation of motion as it stands.  First of all, we need to specify how we are going to chop up our quasi-continuum of particles into differential pieces to determine the mass parcels and the pressure differences across them.  In doing so, we will be able to describe the equation of motion in terms of the variables $\{\rho(x,t),v(x,t)\}$ or, alternately, $\{x(u,t),v(u,t)\}$.  The former will be the corrected Eulerian form of hydro, or x-formulation for short, and the other, the u-formulation.  Notice that the former is the only one that has a nonconstant ``density.''  The u-formulation (index formulation) lets the location and velocity of the particles evolve on top of the fixed index space in contrast with the density and velocity on the position space.  

\subsubsection{u-formulation}
Consider the particle number $N$ of net mass $M$ to be finite and the $u$-parameter space to be of length $L$. We partition the space up into uniform segments $\Delta u$ so that $\Delta m= \frac{M}{L} \Delta u$ for all time.  The RHS of eqn.\ref{eom} in terms of the variable $u$ is $\Delta P(s(u),t)$ but pressure is defined by the $x$-density $P(x)=\beta n(x)$.  Since $n(x(u))=\alpha (ds(u)/du)^{-1}$ and $\alpha L=N$ we have
\begin{align}
	\frac{M}{L}\partial_t^2 s(u,t)&=-\alpha\beta\frac{\partial}{\partial u}\frac{1}{\partial_u s(u,t)}\\
	&=\alpha\beta \frac{\partial_u^2 s(u,t)}{(\partial_u s(u,t))^2}
\end{align}
Using that $M/N=m$, the mass of a particle we have
\begin{align}\label{eomu}
	m \partial_t^2 s(u,t)=\beta \frac{\partial_u^2 s(u,t)}{(\partial_u s(u,t))^2}
\end{align}
Unfortunately, this is defined on the particle index space and so is not very intuitive and hard to relate to actual physical data.  In contrast, the $x$-formulation approach will be defined in terms of variables on real space but with a more peculiar type of pde for the solution.  One possible advantage of it is that it is only two simple terms and for higher dimensional versions may prove a more advantageous route to show evolution is well defined for all time.  The Millenium Challenge problems include a proof of such long term behavior of N-S. However, we will find that N-S is not exactly correct for higher dimensional cases.  Eqn.\ref{eomu} takes the form of a wave equation with variable function dependent $u$-wavespeed 
\begin{align}
	c=\frac{\sqrt{\beta/m}}{\partial_u s}
\end{align}
Since the initial data $s(u)$ must be a one-to-one function with positive slope we see that the rate of propagation of changes grows ever faster as the slope flattens.

\subsubsection{x-formulation}
To attempt to rewrite these equations in terms of the $x$-formulation we begin by 
observing the velocity field is 
\begin{align}
v(u,t)=\frac{\partial}{\partial t}s(u,t)
\end{align}
we can approximate eqn.\ref{eom} arbitrarily well with decreasing $\Delta x$ as 
\begin{align}
\Delta m\cdot\frac{\partial}{\partial t}v(u(x,t),t)=- \beta \cdot\Delta x\cdot \frac{\partial}{\partial x} n(x,t)
\end{align}
where $u(x)=s^{-1}(x)$ $\forall t$.  The resulting equation is
\begin{align}\label{eomx}
\rho(x,t)\frac{\partial}{\partial t}v(u,t)\Big\vert_{u=s^{-1}(x)}=- \beta \frac{\partial}{\partial x} n(x,t)
\end{align}
where $\rho(x,t)=\gamma n(x,t)$.  It is unfortunate that we cannot simply replace $v(u,t)\rightarrow v(u(x,t),t)=v(x,t)$ in eqn.\ref{eomx} since that would give us a way to quickly convert our equation to the new variables $\{\rho(x,t),v(x,t)\}$.  As it stands eqn.\ref{eomx} is not really an eom in the required variables and there is no clear path to manipulate it into them.  

The approach I will use is to take data from the $x$-formulation, pull it back into the u-formulation, evolve by the pde eqn.\ref{eomu}, then push it forwards to $x$-formulation data again as in the following diagram:
\[
\begin{tikzcd}
U(t=t_0) \arrow{r}{\mathcal{O}_u}  & U(t=t_1) \arrow{d}{\mathcal{T}} \\
X(t=t_0) \arrow{u}{\mathcal{T}^{-1}} \arrow[dotted]{r}{\mathcal{O}_x} & X(t=t_1)
\end{tikzcd}
\]

The operator $\mathcal{O}_u$ is the finite time propagation of the data by the pde in eqn.\ref{eomu}.  Our pde is merely the infinitesimal statement of such propagation.  We are especially interested in the form of the analogous infinitesimal propagation for $\mathcal{O}_x$.  

The actions of $\mathcal{T}$ are given by 
\begin{align}
	\mathcal{T}_{(1)}:& s(u)\rightarrow \alpha \frac{ 1}{\partial_u s(u)}\evaluated_{u=s^{-1}(x)}=n(x)\\
	\mathcal{T}_{(2)}:& \{s(u),v(u)\}\rightarrow v(s^{-1}(x))=w(x)
\end{align} 
which time independent. I have used $w(x)$ to indicate the velocity in terms of x to avoid confusion as to which function has which argument.  

The inverse actions are complicated by the fact we need to integrate and have an integration constant that changes in time 
\begin{align}
\mathcal{T}^{-1}_{(1)}:& \{n(x,t),w(x,t),t\}\rightarrow \hat{\mathcal{I}}\small\circ\int_{s_{0}(t)}^{x}\frac{n(x',t)}{\alpha}dx'=s(u,t)\\
\mathcal{T}^{-1}_{(2)}:& \{n(x,t),w(x,t),t\}\rightarrow w(s(u,t))=v(u,t)
\end{align}
where $s_0(t)$ is the location of the $u=0$ particle at time $t$ and $\hat{\mathcal{I}}$ is the function inverse operator where $t$ is treated as constant for its application.  In practice we can define $s(u=0)=0$ at $t=0$ but for later times this position changes and depends on the dynamics of the system in a rather complicated manner.  If we confine ourselves to an effectively infinitesimal time increment we can use $s_0(dt)=v(u=x=0)dt$.  

Let us look at evolution over the time interval $[0,\Delta t]$.  At lowest interesting order we look at two orders of approximation of all the relevant quantities i.e.\
\begin{align}
	n(x)&=n_0+n_1 x\\
	u(x)&=\alpha^{-1}(n_0 x+n_1 x^2/2)\\
	s(u)&=\frac{\alpha}{n_0}u-\frac{\alpha^2 n_1}{2 n_0^3}u^2\\
	w(x)&=w_0+w_1 x\\
	v(x)&=v(s(u))=w_0+w_1\left(\frac{\alpha}{n_0}u-\frac{\alpha^2 n_1}{2 n_0^3}u^2\right)
\end{align}
The values of $s(u,0)$ and $v(u,0)$ can be updated by eqn.\ref{eomu} to give
\begin{align}
 	s(x,\Delta t)&\approx w_0 \Delta t+\alpha\left(\frac{1}{n_0}+\frac{ n_1 w_1}{n_0}\right)u\\
 	v(x,\Delta t)&\approx w_0 -\frac{\beta n_1}{m n_0}\Delta t+\alpha\left( \frac{w_1}{n_0}-\frac{2\beta n_1^2 }{m n_0^3}\Delta t\right)u
\end{align}
To form the necessary inversions to finish the steps in the above diagram we expand about the new center $s_0=w_0\Delta t$ and invert to obtain
\begin{align}
	u'(x,\Delta t)&\approx w_0\Delta t-\frac{w_0(\alpha+n_0)}{\alpha}\Delta t+\left(\frac{n_0}{\alpha} +\frac{(\alpha n_1 w_0-n_0^2 w_1)}{\alpha n_0 }\Delta t \right)x
\end{align}
Evaluating we obtain the following results at $x=0$ to order $\Delta t$
\begin{align}
	\frac{\Delta w}{\Delta t}&=-w_0 w_1-\frac{\beta n_1}{m n_0}\\
	\frac{\Delta n}{\Delta t}&=-n_0 w_1 -n_1 w_0
\end{align}
which gives the usual 1-D Navier-Stokes equations for a linear pressure term when the time increments are limited to zero.  When written in usual vector calculus format we have
\begin{align}
	\partial_{t}\vec{w}&=-\vec{w}\cdot \nabla \vec{w}-\nabla P\\
	\partial_t n&=-\nabla(n\vec{w})	
\end{align}

When these calculations are carried out to two higher orders they give exactly the same result which suggests that they are exact to all orders.  This would seem to validate the usual derivations of hydrodynamic equations by conservation laws.  Let us now quickly look at the 3D case.  

\subsection{3D Hydrodynamics}

Let us consider the 3D generalization of eqn.\ref{eomu}.  The density takes the form
\begin{align}
	n(\vec{x})=\alpha |\mathcal{J}(\vec{s},\vec{u})|^{-1}\vert_{\vec{u}=\vec{s}^{-1}(\vec{x})}
\end{align}
where $\mathcal{J}= \partial s/\partial u$ indicates the Jacobian and $||$ is the determinant.  Since particles can never cross each other so as to create inversions of the local map, we can assume that it is positive definite.  The density is now a volume density so that we use $\alpha L^3=N$ and then
Newton's laws guarantee that the eom is
\begin{align}
	m\partial_t^2 {s_i}(\vec{u},t)=&-\beta \nabla_{{u_i}} \frac{1}{|\mathcal{J}|}\label{eomu3d}\\
	&=\beta \text{Tr}\Big[\frac{\text{Cof}(\mathcal{J})^{\dagger}}{|\mathcal{J}|^2}\nabla_{u_i}\mathcal{J}\Big]\\
	&=\beta \frac{[\text{Cof}\mathcal{J}^{\dagger}]_{jk}}{|\mathcal{J}|^2}\nabla_{i}\mathcal{J}_{kj}\\
	&=\beta \frac{[\text{Cof}\mathcal{J}^{\dagger}]_{jk}}{|\mathcal{J}|^2}\frac{\partial^2 s_k}{\partial u_i \partial u_j}
\end{align}
where Cof is the matrix of cofactors built from the Jacobian.  

Deriving the inverse of these equations looks substantially more difficult and I won't pursue it here.  Vorticity is the most important new feature of hydrodynamics that is introduced by passing beyond one dimension and what will become singular should N-S fail to preserve finite data in finite time.  It would be interesting if this formulation encoded that problem in a new useful form.  

\section{Viscosity}
\subsection{Causality}
One of the easiest ways I can demonstrate that finite order equations will typically fail for hydrodynamics in higher dimensions is to consider the case of laminar flow with shear in a 2D gas with flow given by $w(x,y,t)=v_f(x,t)\hat{y}$.  If the gas is at uniform temperature and density, $\rho_0$, then the force density can be given in terms of the thermal velocity $v_{th}$ and mean free path $\lambda$ by a simple transformation of eqn.\ref{diffusioneom}.  By replacing the particle density with the flow velocity $n\rightarrow v_f$, the diffusion velocity with the thermal velocity $v\rightarrow v_{th}$ and step size with mean free path $\Delta x\rightarrow \lambda$
\begin{align}
	\partial_t v_f(x,t)&=v_{th}\left( \frac{1}{2}\lambda\partial_x^2 v_f(x,t) -\frac{5 }{24} \lambda^3 \partial_x^4 v_f(x,t) + \frac{61}{720}\lambda^5  \partial_x^6 v_f(x,t)  + \ldots\right)
\end{align}
One can argue that the equilibration time and distance is slightly different than the thermal collision time and distance so that these parameters should be changed but there is really a distribution of them so this result for momentum diffusion should be modified as the corresponding case of particle diffusion.  Regardless, the finite step processes imply that no finite order pde is going to preserve causal and correct behavior for gaseous fluids.  

Now consider the case of liquid fluids and viscosity.  The momentum transfer in this case is mediated by elastic strain in the liquid.  If the flow change is rapid this could lead to elastic recoil leading to sound waves.  We generally expect this to be a rather small effect although turbulence does create some acoustic output.  If the particle flowlines, $\vec{s}(\vec{u},t)$, are such that the density is uniform for all time and there are no lateral rearrangements, i.e.\ extensions of flowlines, then the particles move in long uniform chains that may bend but only generate lateral shear forces against the neighboring flows with no interparticle exchanges between them.  This will be the starting point for investigating viscous forces in liquids from a microscopic point of view. However, let us first consider the causality problems associated with viscosity of a simple linear shear.  

Consider a 2D velocity field $\vec{v}=( f(y) ,0)$.  N-S is easily derived for such a Newtonian laminar shear as 
\begin{align}
	\partial_t f(y,t)=-\nu \partial_y^2 f(y,t)
\end{align}
which is the usual diffusion equation with infinite velocity of propagation.  This means that, if we wiggle a horizontal plate in the fluid, the effects propagate instantly through the entirety of it.  The now evident resolution is to consider the fluid layer's microscopic finite thickness, $\Delta s$, corresponding to molecular scale, to have a finite velocity of propagation across it (likely less that longitudinal sound speed) and use the finite scales to generate a causal transfer of momentum across them.  This is an infinite order spatial derivative relationship but this and the above observation for gases are not the most profound observations to be made here regarding viscosity and the finiteness of order in the evolution equations.

\subsection{Microscopic Flow Decompositions}
In rheology, we often work with the deviator, $D_{ij}=\partial_{(i}v_{j)}$ then use symmetry and matrix decomposition properties to build up various viscosity models.  Symmetrization removes antisymmetric parts we know should include rotation and tensorial constructions lead to symmetric stress tensors.  Such objects lead easily to objective forces which are independent of translation, rotation and boost of the system.  I will show that, aside from being hard to connect with microscopic understandings of a liquid, this is not the most general description with these properties.  When we speak of ``shear" viscosity we generally lump both the effect of shearing parallel layer and ``extension'' flow, as in the case of a drawn out filament, into this same term.  As I have argued before, these modes of deformation have different effects on the orientation and diffusion of reactants and other orientable components of the liquid which should be kept distinct for these purposes. Unfortunately, pure linear shear and extension are too general to combine into a unique decomposition but a more refined microscopic analysis of liquids will show that important features of these flows inspire a decomposition that is truly unique with microscopically relevance for the action of flows on constituents.  
 It will be shown that this leads to new sorts of infinite order behavior for fluids with such non-Newtonian and microscopically anisotropic character.  

\subsubsection{Shear vs. Extension in Physical Liquids}
The term ``liquid'' gets are variety of uses in physics.  In the quantum theory of many body systems, it merely relates to a system in which particle motion is highly correlated.  This can be due to long quantum wavelength effects but it has never even been shown that these systems behave in a fashion that resembles classical hydro in any serious manner even though it is often used as a template to describe ultracold experimental behavior.  The term is also used to describe classical fluids on one side of a murky region called a crossover from liquids to gases based on density.  In this article, I mean it only to be the case of fluids that are very close to solids in their density where there is always van der waals and Hydrogen bonding between neighbors.  There is strongly correlated motion due to the near constant density with thermal diffusion creating rearrangements much faster than  hydrodynamic stresses.  In fact, the fluidity of the phase itself piggy backs on this thermal migration by biasing it at the microscopic level.  

One of the classical results of hydrodynamics is that a pure shear flow can be written as a linear combination of a pure extension and a rotation.  In 2D this has the form 
\begin{equation}\label{example}
\nabla v = 
\left(\begin{array}{cc} 0 & -1\\ 1 & 0 \end{array}\right)+\left(\begin{array}{cc} 1 & 0\\ 0 & -1 \end{array}\right)=
\left(\begin{array}{cc} 1 & -1\\ 1 & -1 \end{array}\right)=S(\pi/4)\left(\begin{array}{cc} 0 & 0\\ 2 & 0\\
 \end{array}\right)S^{-1}(\pi/4).
\end{equation}
I will now show how physical liquids made of constituent particles must give a real distinction between these two kinds of deformations then return to what is needed in addition to the deviator to arise at a physical basis for viscosity and rheology in general.  

Given a fluid with some concentration gradient shear and extension produce markedly different results.
Since shearing of a flow in parallel sheets, as in Couette flow, does not require any mixing of layers and extension flow, taken to the extreme of completely flattening a parcel then reextending it like a drawn and flattened filament, mixes all layers along the axis of shortening, there is a real physical distinction between these actions.  

Now let us consider the microscopic stresses of particles in a fluid due to these motions and their physical signatures.  In pure shear flow, $v=\alpha y \hat{x}$, there is a traction force parallel to each layer of particle thickness $\Delta s$. 
The particles of size $\sim \Delta s$ in the fluid rotate in the flow with orbital frequency
\begin{align}
\Omega\lesssim \hat{v}\cdot \nabla_n v 
\end{align} 
These generate transverse forces with their neighbors to prevent continued angular acceleration so that $T_{yx}=T_{xy}$.  The net forces on each particle is then diagonally creating an extension along the $\hat{x}-\hat{y}$ axis.  If the constituent or inclusion particle has some preferred axis of spectroscopic excitation it will align with this deformation with period $\Omega$.  In sea of many randomly oriented particles this signal ceases to have any temporal variation but, if these are dilute inclusions with a long axis of extension then they can be oriented by extension flow or external EM field initially and can maintain some collective order over some time scale.

Conversely, consider pure extension flow, $v=\alpha x \hat{x}-\alpha y \hat{y}$.  In this case such particles tend to align with the $\hat{x}$ axis and have no period of rotation.  This shows that there are physical means to determine variations in microscopic physics due to such flows.  Since rheology depends on this, it seems very likely that there should be meaningful distinctions between shear and extension that are necessary for a general rheological response.  Pure rigid body rotation creates no changes in local bonding so should create no viscous stress.  

Pure extension tends to align polymers and other asymmetrical particles with the flow while pure shear generates torques that tend to rotate them.  This suggests that a combination of such flows will create a competition that determines the microscopic results.  This will affect both the viscous shear and spectroscopic data but requires an analysis at less than the continuum limit to disentangle.

\subsubsection{Unique Decomposition}
There is an extensive literature of decompositions of flow in rheology \cite{Bird,Oldroyd,Truesdell}.  The corotational and codeformational formalisms are among the most popular.  In the corotational case, one defines a frame of rotation then appeals to the polar matrix decomposition, $A=UP$, where $A$ is unitary and $P$ is positive definite hermitian.  Intuitively, $A$ is producing a rotation and $P$ produces deformations.  The general assumption is that the local flow about a particle, when boosted to its reference frame, is $D=\nabla v$.  The decomposition is appealing because $D^\dagger D$ cancels the rotational part and leaves only deforming parts of the matrix.  One flaw in this thinking is that $D$ alone is sufficient to describe the actual local flow.  We can view it as generating an infinitesimal transformation of local points but it is hard to say exactly what is the rotation of such a general ball of points under a general linear velocity field. The attitude here seems to be: ``Whatever the rotation is, we just canceled it out.'' Another issue is that it is unclear why a product decomposition of the local flow is interesting.  A linear sum of flow components is a meaningful way to express the various small displacements in a basis fashion.  Below I will demonstrate that the first order information is not sufficient to make this decomposition but it is well defined and that it is actually important to know how much rotation there is.  

Instead of treating parcels of fluid as primary objects, we return to the case of individual constituent particles as in the previous section.  Even though this is a heavily idealized and granular picture of a fluid undergoing rapid small scale thermal disruptions, it does give a nice model to investigate some of the questions: 1.\ ``What is it that determines if the particle is rotating?'' and 2.\ ``Are its neighbors rotating in a rigid body fashion relative to it?''  The following is a 2D discussion for constant density flow but the argument generalizes to 3D with some more work \cite{Chafin}.  

As a first example, consider global flows of rigid body motion and linear shear.  
\begin{align}
\vec{v}_{\text{shear}}=\left(\begin{array}{cc} 0 & \alpha\\ 0 & 0 \end{array}\right) \left(\begin{array}{c} x \\ y \end{array}\right)\\
\vec{v}_{\text{rot}}=\left(\begin{array}{cc} 0 & \Omega\\ -\Omega & 0 \end{array}\right) \left(\begin{array}{c} x \\ y \end{array}\right).
\end{align}
Near the origin these are evidently different but as we move further away and look at the flow at some point say, $(0,1)$ and look at the local flow about this point in some $\epsilon$-neighborhood we find that the two flows are $\vec{v}_s\approx\alpha (1-y) \hat{x}$ and $\vec{v}_r\approx\Omega (1-y)\hat{x}+\Omega\hat{x}$ which shows these flows are locally identical except for the boost.  Of course, as we continue to second order corrections, the differences become more obvious.  The global linear nature of the flow in the variables $(x,y)$ probably has misled many into not noticing that, for rotation, there are higher order corrections for expansions about other points than the origin.  This should be obvious since the flow has curvature.  We are interested in flows where the rotation is occurring while parcels are translating along arced paths so this consideration is rather important.  It shows that, even for stationary flows, the local linear approximations to the flows are not adequate to extract information about rotation.  We shall soon see that, for nonstationary flows, no finite order information in the spatial derivatives of a flow at fixed time is sufficient to extract the rotation of a parcel.  

Since rigid rotation of a finite parcel of fluid produces no internal stresses, it is natural to assume that a fluid deformation will include as much of it in the form of such as possible. 
Part of the reason such a unique decomposition has been overlooked is that it is assumed that we can use spatial derivatives of the flow $\vec{v}(x,t)$ at fixed $t$ to adequately describe these changes.  By the above linear dependence of general linear shear extension and rotation, it has generally been thought that there is no way to pull out a ``maximal'' rotation.  

First note that if we are looking for the local path of a constituent and its neighbors, the velocity field at a single instant in time is not sufficient to derive this result for the reason that pathlines and streamlines are not the same.  We need to track the actual Lagrangian path of a particle and its neighbors to determine if the flow is causing shearing motions between it and its neighbors.  Even the second order information in $\nabla^2 v$ is not adequate for this purpose. We need the details of the path from recent previous times.  Let us term the rotation a particle undergoes moving along a rigid rotator as ``arc rotation.''  As we saw above, this gives a local flow in most frames that appear like a shear.  It is also possible that a parcel will rotate about this path as it translates along the pathline.  The strategy I will follow is to decompose $D$ by using the osculating circle of the pathline (which is acceleration dependent) to remove the component of rotation not evident in $D$ itself. Once this component is removed the rest of the local flow is determined from $D$.  

Consider a particle in a rigidly rotating disk and decompose the acceleration into components parallel and perpendicular to the velocity field $\vec{a}=\vec{a}_\perp+\vec{a}_{||}$.  The local radius of curvature of the osculating circle to the Lagrangian path is 
\begin{align}
r_\perp&=\frac{v^2}{|\vec{a}_\perp|}\\
\vec{a}_\perp&=\vec{a}-(\vec{a}\cdot\hat{v})\hat{v}
\end{align}  
where the direction of $r_\perp$ is parallel to $a_\perp$ and directed from the convex side of the pathline.  
The osculating plane defined by $\vec{v}$ and $\vec{a}$ give us adjacent arcs that must move with a rigid body velocity field to maintain bond coordination with nearest neighbors.  A local unchanging coordination of bonds between neighbors gives the microscopic definition of rigid local rotation that is meaningful i.e.\ contributing nothing to viscous losses.  The corresponding local angular rotation of the parcel is given by the arc of the path with angular rotation
\begin{align}
	\Omega=\frac{|\vec{v}|}{r_\perp(\vec{a})}\sgn(\vec{v}\times\vec{a}) 
\end{align}
where the local first order motion has yet to be included.  
Using this we have to arc rotation matrix
\begin{align}
A=S(\varphi)\left(\begin{array}{cc} 0 & -\Omega \\ 0 & 0 \end{array}\right)S^{-1}(\varphi)
\end{align}
where $\varphi=\angle(\hat{x},\vec{v})$.  Notice that this seems to be in the form of a shear but, as we saw above, it corresponds to the rotation of the local region about a point $r_\perp$ away from it.

\begin{figure}[h]
	\centering
	\includegraphics[width=3in,clip]{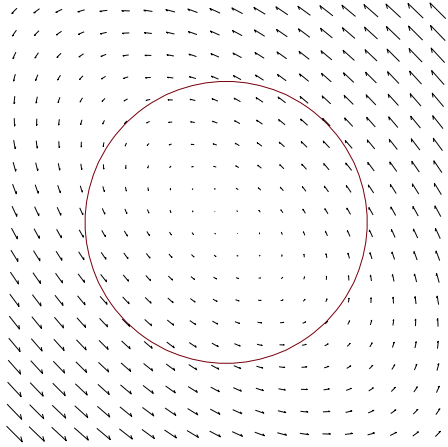}
	\caption{A local multicomponent flow.}
	\label{fig:flow}
\end{figure}

\begin{figure}[h]
	\centering
	\includegraphics[width=4in,clip]{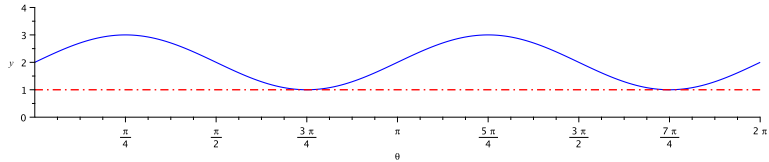}
	\caption{Angular circulation.}
	\label{fig:circulation}
\end{figure}

\begin{figure}[h]
	\centering
	\includegraphics[width=3in,clip]{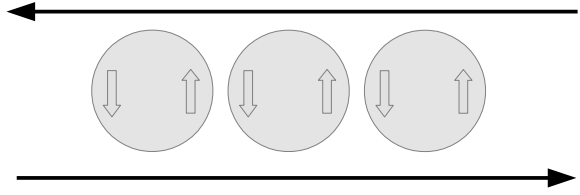}
	\caption{Rolling neighbors in a shear flow.}
	\label{fig:shearballs}
\end{figure}

From here on, the role of neighboring particle motion in generating stresses guide us.  $D-A$ gives a measure of how much uninterupted flow circulates about the particle.  As an example, consider the case of a particle at rest with a local flow given by 
\begin{align}\label{sample}
\vec{v}=\left(\begin{array}{c} -x-2y\\ 2x+y \end{array}\right)
\end{align}
as shown in fig.\ref{fig:flow}
This is a smoothed or averaged version of the microscopically more erratic motion of constituent particles.  We can compute the circulation about the fixed radius ring in the figure (corresponding to the average flow at constituent scale $\Delta s$ in the fluid) as $O(\theta)=\vec{v}\cdot \hat{\theta}\evaluated_{r=1}=\sin(2\theta)+2$ as shown in fig.\ref{fig:circulation}.
Notice that there are no stagnation points on this flow.  This shows that while some neighbors exert more traction than others there is a basic minimal rotating flow rate that all nearest neighbors possess given by $v_{\theta}=\text{Min}|\vec{v}\cdot \hat{\theta}|\text{sgn}(\vec{v}\cdot \hat{\theta})\Delta s=1\cdot\Delta s$ in this case.  The remainder of the flow can be considered pure shear in that there is opposite traction placed on the constituent particle.  The opposite stagnation points given when the minimal rotation is removed corresponds to the axis of rotating neighbors that jam or exert friction to exert canceling torques as in fig.\ref{fig:shearballs}.
The final local decomposition is then 
\begin{align}
\vec{v}=\left(\begin{array}{c} -x-y\\ x+y \end{array}\right)+\left(\begin{array}{c} -y\\ x \end{array}\right)=\vec{v}_s+\vec{v}_r
\end{align}

This gives us an absolute local criterion for rotation for $D-A$, at least for such simple flows.  Once arc rotation is accounted for, the local circulation of particles determine the extent to which they are rotating the particle with not bond disruption versus with a net flow induced torque that has to be canceled by relative rotation with neighboring particles along the pathline.  
Of course some flows, notably extension, have angular stagnation points already present.  In these cases there is no rotation remaining because the flow around the particle is not merely uneven, it is fully interrupted.  I will now classify such flows based on their forces and torques on the central particle and the sort of mixing they induce.

Consider a general $D-A$ matrix and compute the angular circulation at a radius of $\Delta s$ from each point.  The local circulation averages to $C(\theta)\Delta s$ where 
\begin{align}
C(\theta)&= \vec{u}(\theta)\cdot \hat{\theta}\\
\vec{u}(\theta)&=(D-A)\left(\begin{array}{c} \cos(\theta) \\ \sin(\theta) \end{array}\right).
\end{align}
The form of our arc rotation removed matrix is
\begin{align}
	D-A=\left(\begin{array}{cc} a & b\\ c & -a \end{array}\right)
\end{align}
which gives a circulation function of the form
\begin{align}\label{circ}
C(\theta)=H \cos(2\theta+\phi_0)+k
\end{align}
where the relation between the coefficients is given by 
\begin{align}
a&=H \sin(\phi_0)\\
b&=H \cos(\phi_0)+k\\
c&=H\cos(\phi_0)-k
\end{align}
or, inversely,
\begin{align}
H&=\frac{1}{2}\sqrt{4a^2+(b+c)^2}\\
k&=\frac{b-c}{2}\\
\phi_0&=\arctan\left(\frac{2a}{b+c}\right)+\frac{\pi}{4}(1-\sgn(b+c))\sgn(a)
\end{align}
where the last equation gives the full from of the arctan function for $\phi_0\in(-\pi,\pi]$.  This results in a phase shift for the function in $(-\frac{\pi}{2},\frac{\pi}{2}]$ which is fully general since $C(\theta)$ is $\pi$-periodic.  

\begin{figure}[h]
	\centering
	\includegraphics[width=3in,clip]{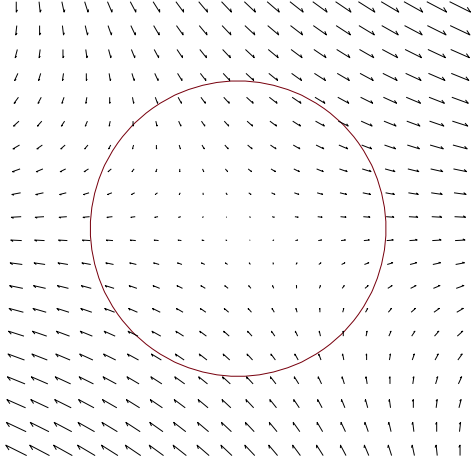}
	\caption{An oblique extension flow.  }
	\label{fig:extflow}
\end{figure}

\begin{figure}[h]
	\centering
	\includegraphics[width=4in,clip]{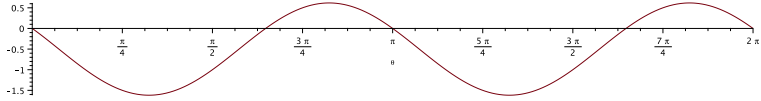}
	\caption{Angular circulation of oblique extension flow.}
	\label{fig:extcirc}
\end{figure}

The form of eqn.\ref{circ} indicates a couple of qualitatively different solutions.  The first is when $|k|>|H|$ which corresponds to the cases like eqn.\ref{sample} where there is continuous flow around the central particle with no stagnation.  The second is the case of $|k|<|H|$ where there are four angular stagnation points as in the case of the local flow eqn.\ref{sample1}
\begin{align}\label{sample1}
\vec{v}=\left(\begin{array}{c} x+y\\ -y \end{array}\right)
\end{align}
The corresponding flow in fig.\ref{fig:extflow} displays a characteristic ``oblique extension'' of such flows where zeros in the circulation in fig.\ref{fig:extcirc} correspond to angular stagnation and extrema correspond to radial stagnation in the flow.  
In this case there is no component of the flow that carries particle around at a uniform rate.  Bonds are being broken on all sides yet there is often a torque still being generated by the flow that must be canceled by some ordered rotation as in fig.\ref{fig:shearballs}.  This torque from the flow is proportional to the circulation
\begin{align}
C_{\tau}= \oint \vec{v}\cdot \hat{\theta} d\theta
\end{align}
even though there is a damping energy loss from breaking of bonds tangent to the circle proportional to
\begin{align}
C_{\kappa}= \oint |\vec{v}\cdot \hat{\theta}|^2 d\theta.
\end{align}

If there is a symmetrical form to $C(\theta)$ then the flow is pure extension.  If not there is an angle in $[0,\pi]$ where
\begin{align}
	\text{Max}|C(\theta)|=\theta_S= \begin{cases}
	\phi_0   & \text{if } k \geq 0 \\
	 \phi_0 +\frac{\pi}{2}      & \text{if } k<0
	\end{cases}
\end{align}
This asymmetry corresponds to a local shear flow described \footnote{The factor of 2 here comes from the circulation of a linear shear of $\alpha=1$ corresponding to $\frac{1}{2}$.}
\begin{align}
T_e=S(\theta_S)\left(\begin{array}{cc} 0 & 2 C_\tau \\ 0 & 0 \end{array}\right)S^{-1}(\theta_S)\sgn(C(\theta_S)).
\end{align}
The remaining flow is a pure extension which preserves radial stagnation points
\begin{align}
	E=D-A-T_e.
\end{align}
For reasons described in sec.\ref{mixing} this is termed the ``interleaving extension'' flow.

In contrast when $|k|>|H|$ the circulation is simply
\begin{align}
R=\frac{1}{2\pi}\left(\begin{array}{cc} 0 & -C_\tau \\ C_{\tau} & 0 \end{array}\right).
\end{align}
so that there is a remaining extension to generate flow driven toques on the central particle given by
\begin{align}
T_s=D-A-R.
\end{align}
Because of its net torque on the local particles in the flow it is termed ``torque shear''.  

Having such a local decomposition we can investigate flow driven mixing as in sec.\ref{mixing}.  More generally we can utilize this decomposition to describe 1.\ the orientation of extended particles in a flow and how they can induce viscous changes, 2.\ spectroscopic effects and alter the chemical reactivity of some species and sites on them and, 3.\ write the stress-tensor of the fluid in a way that each of these independently depends on each torque shear (TS) or interleaving extension (IE) component of the flow as shown next.

\subsection{Stress Tensor and Equations of Motion}
Even though there is no hope of recovering causality at this level of finite approximation, these decompositions allow a very organic decomposition of the stress tensor and, by there dependence on acceleration, will make clear that even this approximated continuum model will not give results that can be expressed in terms of finite numbers of spatial derivatives.  

The local flow decompositions have the forms $D=A+T_e+E$ or $D=A+R+T_s$ where $A$ and hence all the following flows have some dependence on the actual acceleration of the local parcels.  The components of $A$ and $R$ correspond to rotation so have no contribution to viscosity so we can build symmetrized matrices from the other components to give stress components
\begin{align}
	t_{ij}=\begin{cases}
	\eta_{{TS}}T_{e,(ij)}+\eta_{IE}E_{(i,j)}       & \text{if } |k|<|H| \\
	 \eta_{TS}T_{s,(ij)}      & \text{otherwise}
	\end{cases}
\end{align}
This gives an equation of motion for our constant density flow as
\begin{align}
	\partial_t \vec{v}+\vec{v}\cdot\nabla \vec{v}=-\nabla P + \nabla\cdot t
\end{align}
with $P$ defined by constraint, as usual, to preserve the uniform density of the flow. The torque shear and interleaving extension are assigned different values of viscosity for convenience but they are equal for simple fluids.  

It is important now to reflect on the nature of the equations we have derived.  The acceleration, $\partial_t \vec{v}=\vec{a}$, appears both on the left and is explicit and implicit in the definition of the dissipative stress $t_{ij}$ which further involves a nonanalytic partitioning of the flow into two disjoint classes.  This suggests that there may be unusual impulses in the flow along boundaries which separate otherwise smooth transitions between the two classes of flows. 

In attempting to solve the above equations for $\vec{a}$ we are forced into an iterative solution of ever increasing order.  To understand this, recount the Abraham-Lorentz equation and the radiation of charged particles.  The higher order ``jerk'' term is not to be interpreted as a dynamical variable. The A-L equation is actually an identity among deriviatives and not a true equation of motion since it violates the second order limit on Newtonian kinematics \cite{Rohrlich,Yaghjian}.  The solution is found by iteration of the result that keeps it second order in time and introduces every higher spatial derivative of the fields.  As such, it is a one particle analog to the diffusive, mixing and hydrodynamic results in the paper. 
In general, given any equation of the form $\vec{a}=F(\nabla\vec{v},\vec{v},\vec{a})$, a truncation of the iterative solution means that $\vec{a}$ was only a linear and undifferentiated factor in part of the expression on the left side.  Since this is not the case with our expression, termination is generally impossible.  

\section{Rheochemistry}
Polymer melts and solutions are the most extreme cases of extended reactant where the relative orientation of reactants at the moment of near proximity is important for aligning active sites to react.  Many complicated molecules have flexibility that can make some sites available or occluded for reaction.  These larger molecules have slow enough relaxation times to be substantially affected by flows as evident by shear enhanced and reduced drag and polarization measurements.  This makes it clear that flow should be able to affect chemical reactions but how to relate the local flow to the microscopic behavior has been murky with current models and molecular dynamics does not have the scalability to reach the number of particles to mimic bulk flows.  

Chemorheology is the study of how changes in viscosity give us information about how a reaction is progressing.  I suggest the term rheochemistry as a term for using rheological flow to 1.\ reshape and orient reactants relative to each other 2.\ make active sites available or not and 3.\ alter solvation shells and the extended order around them for the purpose of enhancing desired yields and inhibiting unfavorable ones.  

Having a microscopically meaningful way to decompose flows is essential to this end.  In our oblique extension flow, elongated particles will get drawn along the angular stagnation axis that creastes extension and to evacuate positions along the other angular stagnation points.  Creating an extended state that has fixed orientation also limits the diffusive possibilities and reptation of long polymers.  This suggests such flows can delay the time for reactants to meet adding another control to the reaction.  UV catalysis is often polarization dependent but there is generally no corresponding orientation in a solution.  Such extension flows provide an opportunity to optimize the polarization of EM radiation and that of extended reactants so as to favor some types of bonding and activation energy reduction over others.  

The decomposition given in the previous section was a rather elementary one based on a solvent that was rather isotropic in its extent.  In a polymer melt this is not true and new effects that couple the liquid structure to the flow may arise that give richer and more interesting decompositions.  It is my hope that people will see such a foundation as a promising way to develop the subject and provide a stronger theoretical basis for the interplay of reactions and rheology and to generally enrich our toolbox for synthesis.

\section{Conclusions}
Most of us in physics have been happily indoctrinated with the notion that we are getting closer to truth when the equations become simple or there is a simple shortcut around the difficult direct approach using conserved quantities and symmetry.  Here I have presented a few examples where this is certainly not true.  Some clever work may illuminate more about the complicated universal equation in eqn.\ref{diffusioneom} but it is still far more complex than any of the equations of the hyperbolic or relativistic extensions of diffusion designed to reestablish causality (at the cost of extra unphysical degrees of freedom).  For those of us who prefer the mathematical tools of invariants and PDEs, this may be disconcerting but there have been some aesthetically appealing results along the way.  An example is the role of the $\sech$ function of a differential operator in providing a universal approach to diffusive behavior.  Furthermore, this is a nascent topic.  There may be substantial mathematical architecture and simplifications ready to spring from it.  For some of us, beauty and mathematics is not so important as having a clean connection of the macroscopic behavior to the microscopic dynamics.  The results herein also offer some nose thumbing satisfaction for we who have felt abused by the ever condensing and overly terse explanations in physics and were unconvinced that such arguments were valid.  

I have shown that diffusion, flow driven mixing and viscous hydrodynamics are generally not describable with PDEs and that the finite scales of these systems have persistent effects not erasable by some ``continuum limit.''  By abandoning a very comfortable mathematical edifice we gain a great deal, a clear connection of microscopic to macroscopic effects and causality.  The subtle role of ``arc curvature'' in determining the rotation of a parcel is important in another context.  The Cauchy hypothesis in the elasticity of solids says that the stress force on each point can be found by adding up the resultant stress vectors on any small polyhedral parcel.  In short, the curvature of the faces does not matter.  We can view that as true for this formulation of viscous hydro by the formulation in terms of a stress tensor but it is a rather circularly defined object.  The arc rotation defines a preferred curved surface to establish rotation.  As such, curvature seems to be of some importance.  Also the stresses on parcel faces have no direct information about the oblique extension stagnation axes and their role in preferential orientation, which can have an effect back on the viscosity itself.  It would seem that this point deserves some more investigation.

If there were some simple moral to this paper it would be that a partial differential equation with the highest time derivative so included that it cannot be solved for linearly (or as a simple root) leads to equations of motion that are infinite order in spatial derivatives when we force such a solution.  This relatively simple statement has huge implications, opens the door to more complete descriptions of continuum mechanics, and prevents paradoxical behavior that leads to misguided attempts to correct them within an insufficient framework.  
The work here has almost all been in the 1D or 2D domains.  I did some work on 3D in a previous paper \cite{Chafin} but it did not incorporate some of these ideas that have matured here.  It seems that, not only will the equations need some updating but there may be some significantly new behavior in 3D due to the possibility of extended bodies escaping shear forces by aligning themselves perpendicularly to it.  Flow driven stresses can put potential energy bounds on some thermal motions and orientations so thermodynamic and Brownian effects also should be investigated in this context.

\section*{Acknowledgements}
Much of this work was done while a graduate student at North Carolina State University a few years ago.  I have moved on to entrepreneurial endeavors while continuing this and related work so am no longer formally associated with the University but would like to thank my adviser Thomas Schaefer of NCSU and Greg Forrest and Richard McLaughlin of UNC-CH for interesting discussions.


\begin{thebibliography}{1}
	
	\bibitem{Batchelor} 
	Batchelor GK. In press. Conventions and Notation. In An Introduction to Fluid Dynamics. Cambridge University Press. (doi:10.1017/cbo9780511800955)
	
	\bibitem{Bird} 
	Bird RB,  Curtiss CF,  Armstrong RC, Hassager, O :Dynamics of polymeric liquids- Vol. 2:Kinetic Theory, 2. Ausgabe. 437 Seiten, Preis: £ 59.65. John Wiley + Sons, New York, Brisbane, Toronto, Singapore 1987. Berichte der Bunsengesellschaft für physikalische Chemie 91, 1397–1398. (doi:10.1002/bbpc.19870911222)
	
	\bibitem{Chafin} Chafin CE, Hidden Invariants in Rheology: The Persistent Granular Nature of Liquids, arXiv:1405.0649v2 [physics.flu-dyn], (2014).
	
	\bibitem{Molina} Lopez Molina JA, Rivera MJ, Berjano E. 2014 Fourier, hyperbolic and relativistic heat transfer equations: a comparative analytical study. Proceedings of the Royal Society A: Mathematical, Physical and Engineering Sciences 470, 20140547–20140547. (doi:10.1098/rspa.2014.0547)

	\bibitem{Oldroyd}
	Oldroyd JG. 1950 On the Formulation of Rheological Equations of State. Proceedings of the Royal Society A: Mathematical, Physical and Engineering Sciences 200, 523–541. (doi:10.1098/rspa.1950.0035)
	
	\bibitem{Rohrlich}
	Rohrlich F. 2008 Dynamics of a charged particle. Physical Review E 77. (doi:10.1103/physreve.77.046609)
	
	\bibitem{Truesdell} Truesdell C, Noll W. 1992 The Non-Linear Field Theories of Mechanics. Springer Berlin Heidelberg. (doi:10.1007/978-3-662-13183-1)
	
	\bibitem{Yaghjian}
	Yaghjian A 2006 Relativistic Dynamics of a Charged Sphere. Springer New York. (doi:10.1007/b98846)
\end{thebibliography}
\end{document}